\documentclass[letterpaper,twocolumn,prl,aps,superscriptaddress,amsmath,amssymb,floatfix]{revtex4-1}
\usepackage{mathptmx}
\usepackage[latin9]{inputenc}
\usepackage{color}
\usepackage{amsmath}
\usepackage{amssymb}
\usepackage{graphicx}
\usepackage{esint}
\usepackage[unicode=true,
 bookmarks=true,bookmarksnumbered=false,bookmarksopen=false,
 breaklinks=false,pdfborder={0 0 1},backref=false,colorlinks=true]
 {hyperref}
\hypersetup{
 linkcolor=magenta,urlcolor=blue,citecolor=blue,pdfstartview={FitH},hyperfootnotes=false}

\makeatletter

\usepackage{textcomp}
\usepackage{epstopdf}
\usepackage{amsfonts}

\pdfpageheight\paperheight
\pdfpagewidth\paperwidth

\@ifundefined{textcolor}{}{%
 \definecolor{BLACK}{gray}{0}
 \definecolor{WHITE}{gray}{1}
 \definecolor{RED}{rgb}{1,0,0}
 \definecolor{GREEN}{rgb}{0,1,0}
 \definecolor{BLUE}{rgb}{0,0,1}
 \definecolor{CYAN}{cmyk}{1,0,0,0}
 \definecolor{MAGENTA}{cmyk}{0,1,0,0}
 \definecolor{YELLOW}{cmyk}{0,0,1,0}
}

\usepackage{xcolor}\usepackage{soul}
\newcommand{\ket}[1]{\ensuremath{\left|#1\right\rangle}}

\definecolor{blue}{rgb}{0,0,1}
\definecolor{red}{rgb}{0,0,0}
\definecolor{green}{rgb}{0,0,0}
\newcommand{\red}[1]{\textcolor{red}{ #1}}

\makeatother

\begin{document}
\title{Quantum-enhanced radiometry via approximate quantum error correction}
\author{W.~Wang}
\thanks{These three authors contributed equally to this work.}
\affiliation{Center for Quantum Information, Institute for Interdisciplinary Information Sciences, Tsinghua University, Beijing 100084, China}

\author{Z.-J.~Chen}
\thanks{These three authors contributed equally to this work.}
\affiliation{Key Laboratory of Quantum Information, CAS, University of Science and Technology of China, Hefei, Anhui 230026, P. R. China}

\author{X.~Liu}
\thanks{These three authors contributed equally to this work.}
\affiliation{Center for Quantum Information, Institute for Interdisciplinary Information Sciences, Tsinghua University, Beijing 100084, China}

\author{W.~Cai}
\affiliation{Center for Quantum Information, Institute for Interdisciplinary Information Sciences, Tsinghua University, Beijing 100084, China}

\author{Y.~Ma}
\affiliation{Center for Quantum Information, Institute for Interdisciplinary Information Sciences, Tsinghua University, Beijing 100084, China}

\author{X.~Mu}
\affiliation{Center for Quantum Information, Institute for Interdisciplinary Information Sciences, Tsinghua University, Beijing 100084, China}

\author{L.~Hu}
\affiliation{Center for Quantum Information, Institute for Interdisciplinary Information Sciences, Tsinghua University, Beijing 100084, China}

\author{Y.~Xu}
\affiliation{Center for Quantum Information, Institute for Interdisciplinary Information Sciences, Tsinghua University, Beijing 100084, China}

\author{H.~Wang}
\affiliation{Center for Quantum Information, Institute for Interdisciplinary Information Sciences, Tsinghua University, Beijing 100084, China}

\author{Y.~P.~Song}
\affiliation{Center for Quantum Information, Institute for Interdisciplinary Information Sciences, Tsinghua University, Beijing 100084, China}

\author{X.-B.~Zou}
\affiliation{Key Laboratory of Quantum Information, CAS, University of Science and Technology of China, Hefei, Anhui 230026, P. R. China}

\author{C.-L.~Zou}
\email{clzou321@ustc.edu.cn}
\affiliation{Key Laboratory of Quantum Information, CAS, University of Science and Technology of China, Hefei, Anhui 230026, P. R. China}

\author{L.~Sun}
\email{luyansun@tsinghua.edu.cn}
\affiliation{Center for Quantum Information, Institute for Interdisciplinary Information Sciences, Tsinghua University, Beijing 100084, China}


\begin{abstract}
By exploiting the exotic quantum states of a probe,
it is possible to realize efficient sensors that are attractive for
practical metrology applications and fundamental studies~\cite{Giovannetti2004,Schnabel2010,Giovannetti2011,Degen2017,Pirandola2018}.
Similar to other quantum technologies, quantum sensing is suffering
from noises and thus the experimental developments are hindered~\cite{Huelga1997,Escher2011,Demkowicz-Dobrzanski2012}.
Although theoretical schemes based on quantum error correction (QEC)
have been proposed to combat noises~\cite{Arrad2014PRL,Dur2014PRL,Kessler2014PRL,Demkowicz2017PRX,Sekatski2017,Reiter2017,Zhou2018},
their demonstrations are prevented by the stringent experimental
requirements, such as perfect quantum operations and the orthogonal
condition between the sensing interaction Hamiltonian and \red{the noise Lindbladians}~\cite{Zhou2018}. \red{Here, we report an experimental demonstration of a quantum enhancement in sensing with a bosonic probe with different encodings, by exploring the large Hilbert space of the bosonic mode and developing both the approximate QEC and the quantum jump tracking approaches.} In a practical radiometry scenario, we attain a 5.3~dB enhancement of sensitivity, which reaches $\boldsymbol{9.1\times10^{-4}\,\mathrm{Hz}^{-1/2}}$
when measuring the excitation population of a receiver mode. Our results
demonstrate the potential of quantum sensing with near-term quantum
technologies, not only shedding new light on the quantum advantage
of sensing by revealing its difference from other quantum applications,
but also stimulating further efforts on bosonic quantum technologies.
\end{abstract}

\maketitle

The large Hilbert space of a quantum system and the quantum
superposition principle offer the potential advantages of quantum
physics in many applications and laid the foundation
of quantum information science~\cite{Nielsen}. Recently, supported
by the established quantum state engineering and control techniques,
quantum sensing emerges as one of the most promising near-term applications
to achieve quantum advantage and has attracted tremendous attentions~\cite{Giovannetti2004,Schnabel2010,Giovannetti2011,Degen2017,Pirandola2018}.
For a sensor interrogation Hamiltonian $H_{\mathrm{int}}$, the intriguing
Greenberger--Horne--Zeilinger entanglement states of a collection
of spins~\cite{Leibfried2004} or the Schr\"{o}inger-cat-like states
in a large Hilbert space~\cite{Facon2016,Dietsche2019,McCormick2019,WangNC2019Heisenberg}
could enhance the sensitivity of a quantum probe, because they provide
a large variance of energy $\left\langle H_{\mathrm{int}}^{2}\right\rangle -\left\langle H_{\mathrm{int}}\right\rangle ^{2}\propto N^{2}$
with $N$ being the number of excitations. However, these exotic quantum
states are also more prone to environmental noises, and thus the coherence
times are reduced and the ultimate sensing sensitivity is hard to be
enhanced~\cite{Huelga1997}.

\red{Fortunately}, the large Hilbert space also offers the redundancy for realizing
quantum error correction (QEC) that protects quantum states from decoherences
and imperfections. It has been expected that the Heisenberg limit,
i.e. the sensitivity of the sensing scales with the measurement time
($T$) and excitation number as $\propto\left(NT\right)^{-1}$, could
be achieved by protecting the exotic quantum states via QEC~\cite{Arrad2014PRL,Dur2014PRL,Kessler2014PRL,Demkowicz2017PRX,Sekatski2017,Reiter2017,Zhou2018},
in sharp contrast to the standard quantum limit $\propto\left(NT\right)^{-1/2}$.
However, these QEC-based quantum sensing schemes demand stringent
conditions in experiments. On one hand, the non-local QEC operations
are challenging for multi-qubit systems~\cite{Schindler2011,Reed3,Andersen2020}.
Although a pioneering experiment proves the principle of QEC-enhanced
sensing by prolonging the coherence time of a single electron spin~\cite{Unden2016PRL}, the extension to a larger Hilbert space is in absence. On the other hand, the theoretical assumptions of perfect ancilla or error-free
quantum operations are impractical, and the noises of experimental
systems could not meet the orthogonality requirement in general~\cite{Zhou2018}.
The unraveled fact that the Heisenberg limit could not be practically
attainable~\cite{Zhou2020,Shettell2021} discourages further experimental
exploration of quantum-enhanced sensing via QEC.

\begin{figure}
\includegraphics{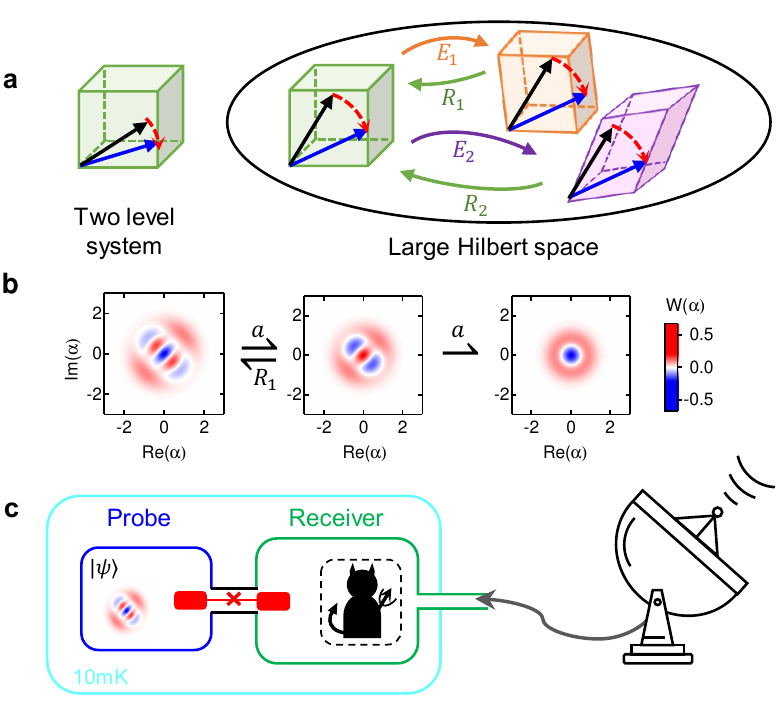} \caption{\textbf{Schematic of practical quantum-enhanced sensing.} \textbf{a}, The principle of quantum sensing with approximate quantum error correction (QEC). The errors $\left\{ E_{1},E_{2},...\right\} $ map the quantum states in the code space to disjoint subspaces and the recovery operations
$\left\{ R_{1},R_{2},...\right\} $ covert the states back to the
code space with the acquired phase being preserved. \textbf{b}, One example
of practical quantum sensing protocol with a bosonic probe. The Wigner
functions illustrate the evolution of the probe quantum state: When the
probe state is initialized to $(|1\rangle+i|3\rangle)/\sqrt{2}$,
the phase can be preserved even if there is a single-photon-loss error ($a$),
which could be tracked and corrected via the recovery operation ($R_{1}$).
\textbf{c}, Schematic of the experimental setup for the quantum radiometry implemented with a superconducting architecture. The device is constructed with a bosonic probe that couples to a receiver mode.}
\label{fig:figure1} \vspace{-6pt}
\end{figure}

In this work, we demonstrate the enhancement of the sensing sensitivity
by approximate QEC with a bosonic probe. Instead of pursuing the Heisenberg
limit, our quantum sensing is implemented with optimized experimental
strategies in a hardware-efficient superconducting architecture. By
using non-exact QEC codes based on two-component Fock states for carrying
the coherence, the sensing information could be protected by QEC,
and the imperfection due to decoherence could even be further
suppressed by mitigating the ambiguity of the quantum evolution trajectories.
Benefiting from both the enhanced sensing-information-gain rate and prolonged coherence time of the exotic quantum states, the bosonic probe is applied for practical radiometry and achieves a detection limit of the receiver
excitation population of $9.1\times10^{-4}\,\mathrm{Hz}^{-1/2}$,
which shows a $5.3\,\mathrm{dB}$ enhancement of the sensitivity by QEC compared to the encoding with the two lowest Fock states. Our
work develops practical quantum sensing technologies, proves the quantum enhancement, and could also stimulate further experimental
efforts in this direction.

\begin{figure*}
\includegraphics{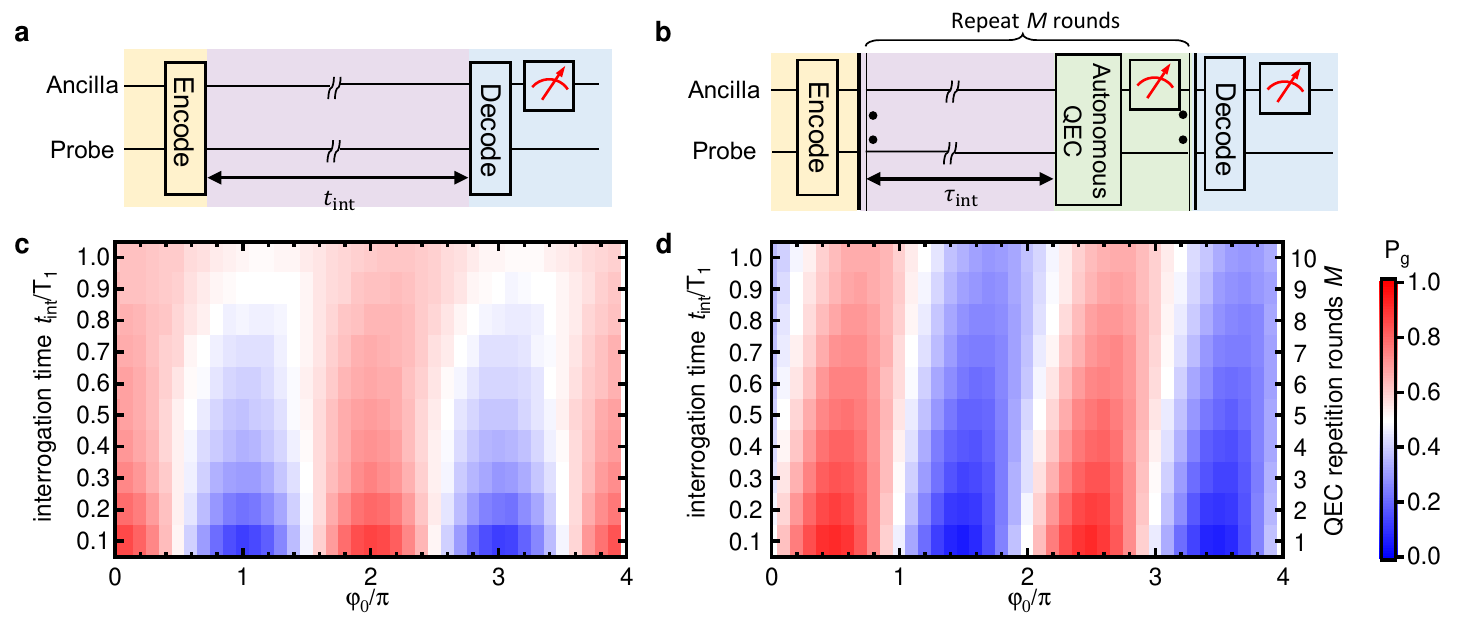} \caption{\textbf{The sensing scheme with QEC.} \textbf{a}, Quantum circuit of the sensing scheme without QEC. After the encoding
process, the probe evolves freely for an interrogation time $t_{\mathrm{int}}$,
and \red{is read out through a detection of the ancilla following a decoding process.} \textbf{b}, Scheme for the QEC-enhanced sensing with the interrogation
interleaved with QEC operations. \red{The QEC operation consists of a QEC pulse and a detection and a conditional reset of the ancilla.} The interrogation time is $t_{\mathrm{int}}$ in \textbf{a} and $t_{\mathrm{int}}=M\tau_{\mathrm{int}}$ in \textbf{b}, with $M$ from 1 to 10 being the repetition number of QEC. \textbf{c}, Measurement results of $P_g$ without QEC for the probe state $\ket{\psi_{1,3}}=(|1\rangle+e^{i\varphi_{0}}|3\rangle)/\sqrt{2}$
against the initial phase $\varphi_{0}$ and \red{with $t_{\mathrm{int}}$
ranging from $0.1T_{1}$ to $T_{1}$.} \textbf{d}, Performance of the QEC-enhanced sensing for the same probe state with an optimized interval $\tau_{\mathrm{int}}=0.1T_{1}$. The shift of fringes with
the increasing $M$ is mainly caused by the QEC-induced phase.}
\label{fig:figure2} \vspace{-6pt}
\end{figure*}

Figure~\ref{fig:figure1} illustrates the principle of the quantum-enhanced
sensing. When a single two-level system (TLS) is used to probe an
external field through coherent coupling, the quantum state in the
two-dimensional Hilbert space would acquire a phase proportional to
the sensing duration $t_{\mathrm{int}}$. By contrast, when extending
the probe to a higher dimension, it is possible to find a subspace
in which the phase accumulation rate is higher, while the rotation
angle can be preserved even though the states are mapped to disjoint
error subspaces by noises. Therefore, the sensitivity could be enhanced
by both the larger rate and the QEC protection. A single bosonic mode
provides an excellent probe for realizing such an idea in practice
because of its large Hilbert space dimension and hardware-efficient
quantum control capability~\cite{Cai2021,McCormick2019,WangNC2019Heisenberg}.
In general, a bosonic probe could sense a physical quantity $\omega$
through the interrogation Hamiltonian
\begin{equation*}
H_{\mathrm{int}}/\hbar=\omega a^{\dagger}a,
\end{equation*}
where $a$ denotes the annihilation operator of the probe and $\hbar$ is the Planck constant.

However, exact QEC codes for sensing with such a model are excluded, since the photon loss errors due to damping are not orthogonal with $H_{\mathrm{int}}$
{[}see Methods{]}. For the convenience of experiments, we propose
the two-component Fock state subspace $\mathrm{span}\left\{ \left|m\right\rangle ,\left|n\right\rangle \right\} $
($m<n$) as the QEC code space for sensing. The evolution of the probe
state after an interrogation time $t_{\mathrm{int}}$ becomes
\begin{equation*}
|\psi_{m,n}\rangle=\alpha_{m,n}|m\rangle+\beta_{m,n}e^{-i(n-m)\omega t_{\mathrm{int}}+i\varphi_{0}}|n\rangle,
\end{equation*}
where $\alpha_{m,n},\beta_{m,n}\in\mathbb{R}$ are the amplitudes
and $\varphi_{0}$ is the initial phase. It is easy to verify that
for loss errors up to $m$ photons, the acquired phase $\varphi=\left(n-m\right)\omega t_{\mathrm{int}}$
is preserved with the phase accumulation rate $\left(n-m\right)\omega$
being fixed irrespective to the time when the photon jump occurs.
By repetitively implementing the recovery operations $R_{j}=\left|m\right\rangle \left\langle m-j\right|+\left|n\right\rangle \left\langle n-j\right|+\widetilde{R}_{j}$ for the $j$-photon-loss error, with $\widetilde{R}_{j}$ being a complementary operator to make $R_{j}$ unitary, the phase coherence could be protected in the code space. Therefore, such a code in a high-dimensional Fock space could simultaneously enhance the phase accumulation rate by $n-m$ times when compared to the encoding with the lowest two levels and prolong the coherence time via QEC. However, the coherence time could only
be enhanced by a limited factor, since the code is an approximate
QEC code and the amplitudes of the probe state $\alpha_{n,m}$ and
$\beta_{n,m}$ change after each photon loss, as illustrated by the
deformation of the error subspaces in Fig.~\ref{fig:figure1}\textbf{a}.

An example of this proposal is schematically shown in Fig.~\ref{fig:figure1}\textbf{b} with a probe state $\ket{\psi_{1,3}}=\frac{1}{\sqrt{2}}\left(|1\rangle+e^{i\pi/2}|3\rangle\right)$,
i.e. $m=1$ and $n=3$. The single-photon-loss error maps the state
to $\frac{1}{2}\left(|0\rangle+\sqrt{3}e^{i\pi/2}|2\rangle\right)$
and recovery $R_{1}$ converts the state back to the code space, while
the Wigner functions of both states have the same rotation
angle and symmetry. However, the phase information is completely corrupted
for a two-photon-loss error since $m<2$.

To demonstrate the efficacy of the approximate
bosonic QEC scheme for sensing, we first experimentally characterize
the performance of the scheme by measuring a virtual phase ($\varphi_{0}$)
introduced in the initial probe state instead of an acquired phase ($\varphi$).
Our experimental device schematically shown in Fig.~\ref{fig:figure1}\textbf{c}
consists of a superconducting transmon qubit as an ancilla dispersively coupled
to two three-dimensional cavities~\cite{Paik,Ofek2016,Hu2018}.
The cavity (blue) with a long coherence time ($T_{1}=143\,\mathrm{\mu s}$)
serves as the probe, while the ancilla and the other short-lived
cavity (green) assist the manipulation and readout of the probe, \red{respectively} {[}see Methods for more details{]}. In the radiometry
experiments studied later, the readout cavity serves as a receiver
to collect microwave signals from outside, while the probe can sense
the excitation in the receiver through the cross-Kerr interaction.

As shown by the experimental circuits in Figs.~\ref{fig:figure2}\textbf{a}
and \ref{fig:figure2}\textbf{b}, the ancilla assists the encoding
and decoding of the probe by mapping the ground and the excited states
$\left\{ \left|g\right\rangle ,\left|e\right\rangle \right\} $ to
the two Fock states $\left\{ \left|m\right\rangle ,\left|n\right\rangle \right\} $.
The sensing procedure is reminiscent of the Ramsey interferometer~\cite{Chu2002}
with the output probability of $\left|g\right\rangle $ as $P_{g}=A+B\cos(\varphi+\varphi_{0})$
manifesting the interference fringe, where $A$ and $B$ are the fitting
parameters. Since the decay rate of Fock state $|m\rangle$ is
proportional to the photon number $m$, we fix $m=1$ and select $n=3,~5,~7$ for a relatively small decay rate. The corresponding
error set of the probe is $\left\{ E_{0}=e^{-t_{\mathrm{int}}a^{\dagger}a/2T_{1}},E_{1}=\sqrt{1-e^{-t_{\mathrm{int}}/T_{1}}}e^{-t_{\mathrm{int}}a^{\dagger}a/2T_{1}}a\right\} $ for zero- and single-photon-loss errors with an interrogation time
$t_{\mathrm{int}}$, and we only tackle the dominant error $E_{1}$
by $R_{1}$. The QEC of the probe state is implemented through an
autonomous manner, i.e. by applying the correcting pulse after an
interrogation time $\tau_{\mathrm{int}}$ to act $R_{1}$ on the probe
and flip the ancilla if $E_{1}$ occurred {[}see Methods{]}. Note
that $\alpha_{m,n}$, $\beta_{m,n}$, and $\tau_{\mathrm{int}}$
are optimized in order to maximize the visibility of the output fringes
for each $\left|\psi_{m,n}\right\rangle $ in the following experiments,
therefore the best detection sensitivity can be achieved experimentally
{[}see Methods{]}.

Figures~\ref{fig:figure2}\textbf{c} and \ref{fig:figure2}\textbf{d}
compare the measured probability $P_{g}$ against the virtual phase
$\varphi_{0}$ and the sensing duration $t_{\mathrm{int}}$ for the
cases without and with the protection by QEC. The results with QEC
indeed show a much slower decaying of the fringe visibility. To evaluate
the potentially achievable improvement of the sensing performance, we extract the Ramsey visibility against $\varphi_{0}$ and derive the normalized quantum Fisher information $\mathcal{Q}$ with respect to $t_{\mathrm{tot}}$, which is the total experimental time for a single-shot measurement including the initialization, encoding, interrogation, decoding,
and readout, as well as the time needed for the QEC process and feedback
when QEC is performed [see Methods]. $\mathcal{Q}$ determines the
best achievable sensitivity of $\omega$ in a unit time, i.e.
\begin{equation}
\sigma_{\omega}\leq1/\sqrt{\mathcal{Q}},
\end{equation}
which is \red{in the unit of $\mathrm{Hz}/\sqrt{\mathrm{Hz}}$} and corresponds to the
\textcolor{red}{noise floor of our sensor}.

\begin{figure*}
\includegraphics{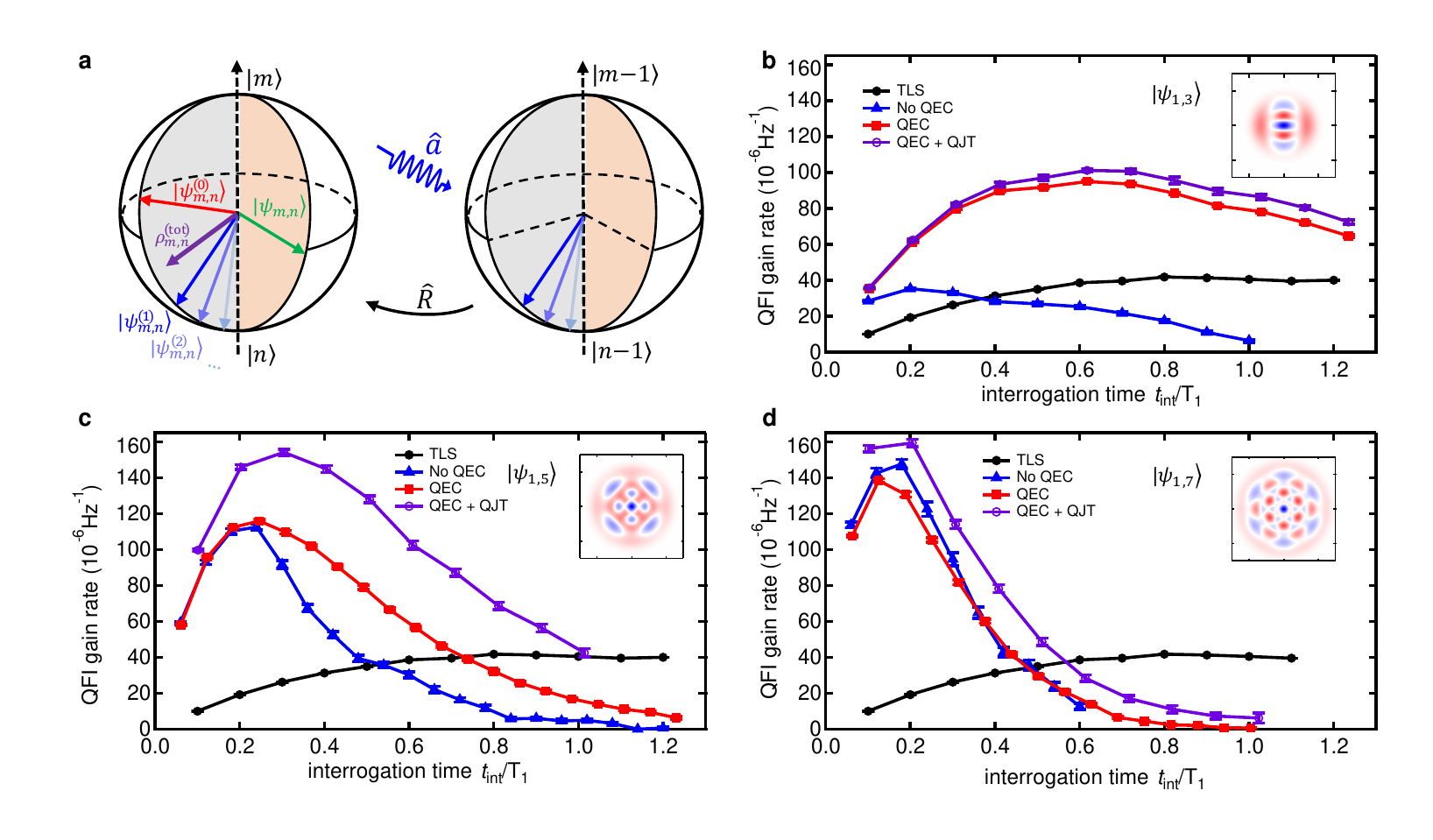} \caption{\textbf{Normalized quantum Fisher information $\mathcal{Q}$.} \textbf{a}, The Bloch-sphere illustration of the quantum jump tracking (QJT). For a two-component Fock state $\left|\psi_{m,n}\right\rangle $, although it could be confined in the code space $\mathrm{span}\left\{ \left|m\right\rangle ,\left|n\right\rangle \right\} $ \red{via QEC}, the amplitudes vary depending on the number ($j$) of the single-photon-loss
errors occurring, i.e. $\left|\psi_{m,n}\right\rangle \protect\mapsto\left|\psi_{m,n}^{(j)}\right\rangle $. The output would be a mixed state $\rho_{m,n}^{\mathrm{(tot)}}$, if different evolution trajectories could not be distinguished. \textbf{b}-\textbf{d}, Quantitative performance of different quantum sensing strategies \red{characterized by $\mathcal{Q}$} for the probe states $\left|\psi_{1,3}\right\rangle$, $\left|\psi_{1,5}\right\rangle $, and $\left|\psi_{1,7}\right\rangle $, respectively. TLS: the two-level
system encoding with the two lowest Fock states. QEC \& No QEC: results
with and without QEC, respectively. QEC+QJT: the strategy that combines QEC and QJT. The error bars are obtained through \red{error propagation of the fit parameter uncertainties.} Inset: Wigner fucntions of the corresponding probe states, with the same axes and color scale bar as in Fig.~\ref{fig:figure1}b.}
\label{fig:figure3} \vspace{-6pt}
\end{figure*}

\begin{figure*}
\includegraphics{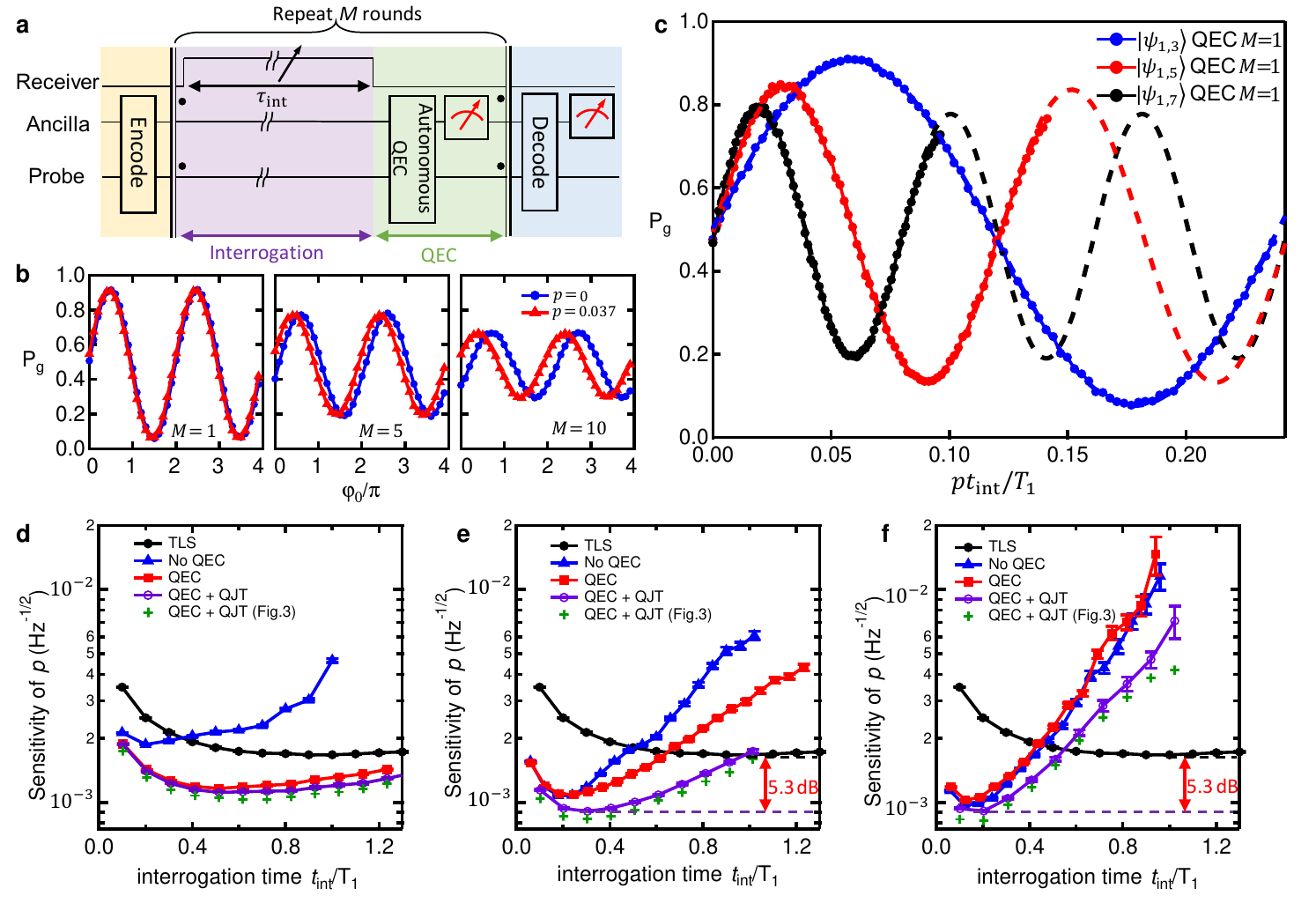} \caption{\textbf{The quantum radiometry.} \textbf{a}, Experimental sequence for the quantum-enhanced radiometry that senses the excitation population $p$ in the receiver cavity \red{(Fig.~\ref{fig:figure1}c) via QEC}. \textbf{b}, The measured $P_{g}$ as a functions of \red{the initial phase} $\varphi_{0}$. The blue dots and red triangles correspond to experiments with $p=0$ and $0.037$, respectively.
The experiment is performed with the probe state $\ket{\psi_{1,3}}=(|1\rangle+|3\rangle)/\sqrt{2}$
and $\tau_{\mathrm{int}}=0.1T_{1}$ for the QEC repetition number $M=1,~5,~10$
(from left to right). \textbf{c}, The measured $P_{g}$ as a function
of $pt_{\mathrm{int}}/T_{1}$ for the probe states $\left|\psi_{1,3}\right\rangle $, $\left|\psi_{1,5}\right\rangle $, and $\left|\psi_{1,7}\right\rangle $ with a single round of QEC ($M=1$ and \red{$t_\mathrm{int}=M\tau_\mathrm{int}=\tau_\mathrm{int}$)}. The fitted \red{oscillation} periods are proportional to $n-m$. \textbf{d}-\textbf{f}, Sensitivity of measuring $p$ ($\sigma_p$) of the radiometry for the probe states $\left|\psi_{1,3}\right\rangle$, $\left|\psi_{1,5}\right\rangle $,
and $\left|\psi_{1,7}\right\rangle $, respectively. A sensitivity enhancement of 5.3~dB over TLS (the encoding with the two lowest Fock states) is obtained. QEC+QJT (Fig.3): the deduced sensitivity from the results with the QEC+QJT strategy in Fig.~\ref{fig:figure3}. The error bars are obtained through error propagation of the fit parameter uncertainties.}
\label{fig:figure4} \vspace{-6pt}
\end{figure*}

The results of $\mathcal{Q}$ with and without QEC are summarized in Figs.~\ref{fig:figure3}\textbf{b-d}, and $\mathcal{Q}$ of the simple TLS case with the probe being encoded in the lowest two levels is also provided as a reference. All curves of $\mathcal{Q}$ \textcolor{black}{first grow up with} $t_{\mathrm{int}}$, \textcolor{black}{since a longer interrogation time gives a larger phase acquisition. However, due to more accumulated decoherence, $\mathcal{Q}$ saturates at an optimal $t_{\mathrm{int}}$ and decreases for even longer $t_{\mathrm{int}}$. Benefiting from the larger Hilbert space dimension, the best achievable $\mathcal{Q}$ without QEC increases with $n$, while the optimal $t_{\mathrm{int}}$ decreases with $n$ due to the shorter Fock state lifetime. When performing QEC, $\mathcal{Q}$ is clearly improved than that without QEC especially when $t_{\mathrm{int}}$ exceeds the optimal value, confirming the enhanced coherence time by QEC. Comparing the results for different $n$, the improvement induced by QEC reduces with increasing $n$, due to the stronger uncorrectable noise effects for larger $n$ and also higher operation errors when performing QEC.}

One main limitation on the performance of the approximation QEC would
be the code state deformation, which causes decoherence even when
the QEC is successfully implemented. For example, after one round
of QEC, the amplitudes of $\ket{\psi_{m,n}}$ evolve as $\left\{ \alpha_{m,n},\beta_{m,n}e^{-\left(n-m\right)\tau_\mathrm{int}/2T_{1}-i\varphi_{0}}\right\} $
and $\left\{ \alpha_{m,n},\sqrt{\frac{n}{m}}\beta_{m,n}e^{-\left(n-m\right)\tau_\mathrm{int}/2T_{1}-i\varphi_{0}}\right\} $
(normalization factors are neglected) for $E_{0}$ and $E_{1}$ occurring,
respectively, with the relative amplitude of the two Fock components
being either amplified or suppressed. As illustrated in Fig.~\ref{fig:figure3}\textbf{a}, the final probe state \red{$\rho_{m,n}^\mathrm{(tot)}$} becomes a mixed state composing of different possible quantum evolution trajectories \red{($\ket{\psi_{m,n}^{(0)}}$, $\ket{\psi_{m,n}^{(1)}}$, ...)}, although the phase is preserved irrespective to the total number of photon jumps during $\tau_{\mathrm{int}}$. If we can distinguish the number of photon jumps that have occurred, the ambiguity of the possible trajectories could be mitigated, and thus $\mathcal{Q}$ could be improved.

Therefore, we propose and demonstrate a quantum jump tracking (QJT)
approach: record the output of each autonomous QEC, count the number
of single-photon jumps ($j\in\left\{ 0,1,...,M\right\} $), and process
the data according to $j$ {[}see Methods{]}. By doing so, $\mathcal{Q}$
is further improved in all cases and even the optimal $t_{\mathrm{int}}$
is extended, as shown in Figs.~\ref{fig:figure3}\textbf{b}-\textbf{d}.
These experimental results demonstrate the protection and recovery
of $\mathcal{Q}$ from decoherence by QEC and QJT, and indicate the
benefits of approximate QEC in sensing.

Finally, the scheme is applied in a practical sensing scenario as
a quantum radiometer. Based on the device shown in Fig.\textbf{~}\ref{fig:figure1}\textbf{c}
and using the readout cavity as a receiver to the microwave field
under detection, the excitation population of the readout cavity $p=\omega/\chi$
could be derived by \red{the quantum sensor} with a calibrated cross-Kerr
coefficient $\chi$ between the probe and the receiver. Following
the sequence shown in Fig.~\ref{fig:figure4}\textbf{a}, the \red{resulting oscillations} shift due to the acquired phase $\varphi$ from the receiver population $p=0.037$ induced by a continuous weak coherent signal {[}Fig~\ref{fig:figure4}\textbf{b}{]}. It is observed
that $\varphi$ increases linearly with the acquisition time $M\tau_{\mathrm{int}}$
as expected, but the contrast of the signal fades due to decoherence and errors that cannot be completely corrected. With an appropriate $\varphi_{0}$, the sensitivity of the output $P_{g}$ could be maximized by optimizing the slope
$\partial P_{g}/\partial p$ for $p\approx0$. The results in Fig.~\ref{fig:figure4}\textbf{c} for different $\left|\psi_{m,n}\right\rangle $
show the maximum slopes around $p\approx0$ proportional to $n-m$.
From these results, the achieved experimental sensitivity for measuring
$p$ is derived as
\begin{equation}
\sigma_{p}=\Delta P_{g}\sqrt{t_{\mathrm{tot}}}/|\partial P_{g}/\partial p|,
\end{equation}
with $\Delta P_{g}=\sqrt{P_{g}\left(1-P_{g}\right)}\approx\frac{1}{2}$
being the standard deviation of $P_{g}$ which follows the binomial
statistics. When QJT is applied, $\sigma_{p}$ could be derived
with $P_{g}$ and its slope conditional on $j$ [see Methods].

Figures~\ref{fig:figure4}\textbf{d}-\textbf{f}
summarize the achieved sensitivities of the quantum radiometry for different
strategies, as well as the potentially achievable $\sigma_{p}=1/\chi\sqrt{\mathcal{Q}}$
based on the virtual phase measurement results in Fig.~\ref{fig:figure3}.
It is confirmed that the best strategy is to combine QEC and QJT and
shows great advantage over others, with the sensitivity $\sigma_{\psi_{1,3}}=11.2$,
$\sigma_{\psi_{1,5}}=9.1$, and $\sigma_{\psi_{1,7}}=9.1$ (in the unit
of $10^{-4}\,\mathrm{Hz}^{-1/2}$) achieved for $\ket{\psi_{1,3}}$, $\ket{\psi_{1,5}}$, and $\ket{\psi_{1,7}}$ respectively. The general trends of the achieved $\sigma_{p}$ agree well with the results deduced from Fig.~\ref{fig:figure3}, but with a slight sensitivity loss due to the $j$-independent decoding \red{in the current experiment} instead of the most optimal adaptive decoding. Compared with the TLS case ($\sigma_{\mathrm{TLS}}=16.7\times10^{-4}\,\mathrm{Hz}^{-1/2}$), we realize a sensitivity enhancement
of $20\log_{10}\sigma_{\mathrm{TLS}}/\sigma_{\psi_{1,7}}=5.3\,\mathrm{dB},$
approaching the optimal enhancement of $6.2\,\mathrm{dB}$ implied
by the results from Fig.~\ref{fig:figure3}.

A demonstration of quantum-enhanced sensing by a bosonic probe is
performed with a superconducting circuit. Utilizing the large Hilbert
space of the bosonic mode, \red{we realize a radiometry that shows
a quantum enhancement of 5.3~dB and opens the door to practical quantum
sensing.} The gain of quantum Fisher information by approximate QEC
and QJT reveals the significant difference between the quantum sensing
and other quantum information processing applications: the goal is
to acquire the sensing information as much as possible instead of
pursuing the perfect protection of an unknown quantum state. The
extensions of the scheme to \red{tens of} photons by developing
sophisticated quantum control method and optimal approximate QEC codes,
as well as to multiple bosonic modes, are appealing and worth further
investigations. The bosonic radiometry demonstrated here will also
excite immediate interests for other quantum sensing applications,
such as the force sensing~\cite{Jacobs2017}, because the scheme is
applicable to all physical quantities that could induce a change of $\omega$. The bosonic probe having the advantages of hardware efficiency and
avoiding the non-local interactions is extensible to other bosonic
degrees of freedom including phonons coupled with trapped ions~\cite{McCormick2019}
and superconducting qubits~\cite{Chu2017}, and also extensible to the collective
excitations in spin and atom ensembles to promote the atomic and optical
quantum metrology technologies~\cite{Giovannetti2004,Schnabel2010,Giovannetti2011,Degen2017,Pirandola2018}.




%

\vbox{}

\noindent \textbf{Acknowledgments}\\
 \textcolor{black}{This work was supported by National Key Research
and Development Program of China (Grant No.2017YFA0304303) and the
National Natural Science Foundation of China (Grant No.11925404 and
11874235). C.-L.Z. was supported by National Natural Science Foundation
of China (Grant NO.12061131011 and No.11874342) and Anhui Initiative
in Quantum Information Technologies (AHY130200).}


\vbox{}

%
%



\cleardoublepage{}

\noindent \textbf{\large{}{}{}{}{}{}Methods}{\large\par}

\noindent \textbf{Experimental implementation}. As described in the
main text, our experiments are implemented with a superconducting
quantum circuit and the device consists of a superconducting transmon
qubit as an ancilla dispersively coupled to two superconducting rectangular
microwave cavities, the probe and the receiver. The input and output
couplings of the receiver cavity are designed to be asymmetric, $\kappa_{\mathrm{r,out}}\gg\kappa_{\mathrm{r,in}}$,
offering a decay rate of $\kappa_{\mathrm{r}}=1/44\,\mathrm{ns}$,
which is three orders of magnitude higher than that of the probe cavity.
As a consequence, a static coherent state with a mean excitation number
$p$ in the receiver cavity can be created within a negligible period
of time. \textcolor{black}{Such a design of coupling is fit for the
high-fidelity readout of the ancilla. However, for practical applications
the receiver mode could be over-coupled to the external fields that are to be detected and an extra readout resonator could be employed to perform the readout.}
Crucially, the cross-Kerr interaction induced by the nonlinearity
of the transmon allows the detection of the excitation in the receiver
by utilizing the bosonic quantum states in the probe cavity.
The calibrated cross-Kerr coefficient is $\chi/2\pi=15.3$~kHz.

\vbox{}

\noindent \textbf{Quantum Fisher Information (QFI).} To characterize
the sensitivity that can be achieved in the experiment, the measured data are fitted with sinusoidal curves. For the simplest
case with binary outputs $\left\{ \left|g\right\rangle ,\left|e\right\rangle \right\} $,
the results obey $P_{g}(\omega)=A+B\cos\bigl[\omega\bigl(n-m\bigr)t_\mathrm{int}+\varphi_{0}\bigr]$
and $P_{e}=1-P_{g}$. The final readout outputs follow the binomial
distribution, giving a variance of $(\Delta P_{g})^{2}=P_{g}(\omega)-P_{g}^{2}(\omega)$.
Therefore, the attainable resolution (uncertainty $\Delta\omega$)
for measuring $\varphi$ by one round of the sensing experiment, including
the initialization, encoding, interrogation, decoding, and readout,
reads
\[
\Delta\omega\leq\mathrm{max}_{\varphi_{0}}\frac{\Delta P_{g}}{\partial P_{g}\left(\omega\right)/\partial\omega}=\frac{\sqrt{A(1-A)}}{\bigl(n-m\bigr)t_\mathrm{int}B}.
\]
Alternatively, the precision could be studied in a more general frame
of quantum metrology, where the QFI gain $\mathcal{F}$ could be deduced
as
\begin{align*}
\mathcal{F} & =\mathrm{max}_{\varphi_{0}}\left\{ \frac{1}{P_{g}(\omega)}\left[\frac{\partial P_{g}\left(\omega\right)}{\partial\omega}\right]^{2}+\frac{1}{P_{e}(\omega)}\left[\frac{\partial P_{e}\left(\omega\right)}{\partial\omega}\right]^{2}\right\} \\
 & =\frac{\bigl(n-m\bigr)^{2}t_\mathrm{int}^{2}B^{2}}{A\left(1-A\right)}.
\end{align*}
The corresponding achievable measurement uncertainty is
\[
\Delta\omega\leq1/\sqrt{\mathcal{F}}.
\]
Introducing the normalized QFI per unit time $\mathcal{Q}=\mathcal{F}/T$,
\red{with $T$ being the experimental time, we then could derive the practical sensitivity of the experiment $1/\sqrt{\mathcal{Q}}$ that is associated with the measurement bandwidth and corresponds to the noise floor of the sensing.} For example, if the duration for a single-shot sensing experiment is $t_{\mathrm{tot}}$ , the QFI gain rate could be obtained as
\[
\mathcal{Q_{\omega}}=\frac{B^{2}\left(n-m\right)^{2}t_\mathrm{int}^{2}}{A(1-A)t_{\mathrm{tot}}},
\]
corresponding to the achievable measurement sensitivity
\[
\sigma_{\omega}\leq\frac{1}{\sqrt{\mathcal{Q}_{\omega}}}.
\]

\vbox{}

\noindent \textbf{Parameter optimization.} For the probe quantum states,
the damping channel could be represented by the Kraus operators as
$\mathcal{E}\left(\rho\right)=\sum_{k=0}^{\infty}E_{k}\rho E_{k}^{\dagger}$,
where $\rho$ is the density matrix of the probe and
\[
E_{k}=\frac{(1-e^{-\frac{t_{\mathrm{int}}}{T_{1}}})^{k/2}}{\sqrt{k!}}e^{-\frac{t_{\mathrm{int}}}{2T_{1}}a^{\dagger}a}a^{k}
\]
is the operator for the $k$-photon-loss error during a sensing interrogation
time of $t_{\mathrm{int}}$. In the experiments, we only consider
the two dominant errors $\left\{ E_{0},~E_{1}\right\} $. Due to the
Fock state damping $e^{-\frac{t_{\mathrm{int}}}{2T_{1}}a^{\dagger}a}$
and \red{the photon-number-dependent photon jump rate},
the amplitudes of the two-component Fock states experience unbalanced
amplitude change. Therefore, we numerically optimize the coefficients
$\alpha_{m,n}$ and $\beta_{m,n}$ of the initial probe quantum states, $\tau_\mathrm{int}$, and the repetition number $M$ to maximize $\mathcal{Q}$ by considering the full damping channel and the imperfections of QEC and other operations. We further optimize the interrogation time $\tau_\mathrm{int}$ experimentally in order to acquire the maximum $\mathcal{Q}$. For the purpose of achieving the maximum sensitivity of the radiometry, we also optimize the initial phase $\varphi_{0}$ in the encoding step to maximize the slope $\partial P_{g}/\partial p$.


\vbox{}

\noindent \textbf{Autonomous implementation of QEC.} The probe quantum
states studied in this work are the approximate QEC codes, and thus the recovery
$R_{j}=\left|m\right\rangle \left\langle m-j\right|+\left|n\right\rangle \left\langle n-j\right|+\widetilde{R}_{j}$
for the $j$-photon-loss error are derived according to the transpose
channel, which provides a universal approach for constructing the
recovery operation with reasonable performance. Such recovery operations
are implemented autonomously and repetitively during the sensing.
This type of autonomous QEC can map \red{the error state} back into the code
space $\mathrm{span}\left\{ \left|m\right\rangle ,\left|n\right\rangle \right\} $
whenever the system jumps to an error state, by sending pulses to
both the probe cavity and the ancilla qubit simultaneously without
the requirement for error detection and real-time adaptive control.
Such a protocol can avoid additional fast feedback electronics and
also circumvent the delay in the electronics, and thus saves the hardware
and suppresses potential decoherence due to the delay. After the autonomous
implementation of QEC, \red{the error entropy is transferred from the probe
state to the ancilla.} Once a single-photon-loss error ($E_{1}$) occurs,
the probe state is recovered and the ancilla is flipped to the
excited state $|e\rangle$. However, when there is no error, both the probe
and the ancilla remain unaltered during the QEC operation. The subsequent
measurement of the ancilla gives a result of $|e\rangle$ or $|g\rangle$
that can be recorded in real time, indicating a single-photon-loss
error occurs or not. A conditional $\pi$-pulse is then applied to the ancilla to
reset it to $|g\rangle$ for the next repetition of QEC. For more information about the autonomous QEC, see Refs.~\cite{Ma_2020,Cai2020}. The whole correction process is repeated for $M$ times followed by a decoding for the final readout to end the sensing experiment.

\vbox{}

\noindent \textbf{Quantum jump tracking (QJT).} The ancilla output
of $\left|e\right\rangle $ after the QEC pulse indicates a single-photon-loss
error occurs during the interrogation time. In the QJT experiments, the number of $\left|e\right\rangle $ at the output of the QEC pulse is counted and recorded, which allows us to improve the sensitivity. In general, the decoding operation should be adaptively selected according to the number of the quantum jumps ($j$). \red{The optimal decoding scheme is to map the Fock states $\ket{m}$ and $\ket{n}$ to the ancilla state $\left|\pm\right\rangle =\left|g\right\rangle \pm e^{\pm i\varphi_{1}}\left|e\right\rangle $ for all cases, with $\varphi_{1}\approx 0$ being the readout phase, therefore we only selectively process the data after the experiments.} We divide the final measurement outputs into groups by $j$. The conditional probability of output $\left|g\right\rangle $, $P_{g,j|M}(\omega)$, for $j$ jumps, corresponding to the case with the single-photon-loss error occurring $j$ times among $M$ repetitions of sensing, can be fitted with
$P_{g,j|M}(\omega)=A_{g,j}+B_{g,j}\cos\left[\omega\bigl(n-m\bigr)t_\mathrm{int}+\varphi_{g,j}\right]$.
Similarly, for output $\left|e\right\rangle $, $P_{e,j|M}(\omega)$ can
be fitted with $P_{e,j|M}(\omega)=A_{e,j}+B_{e,j}\cos\left[\omega\bigl(n-m\bigr)t_\mathrm{int}+\varphi_{e,j}\right]$.
Here, $\varphi_{g(e),j}$ includes the initial and the readout phases.
The resulting normalized QFI $\mathcal{Q}$ with QJT can be
calculated as
\begin{align*}
\mathcal{Q}_{\omega}= & \frac{1}{t_{\mathrm{tot}}}\mathrm{max}_{\varphi_{j}}\left\{ \sum_{j=0}^{M}\sum_{l\in\left\{ g,e\right\} }\frac{1}{P_{l,j|M}(\omega)}\left[\frac{\partial P_{l,j|M}(\omega)}{\partial\omega}\right]^{2}\right\} \\
= & \frac{1}{t_{\mathrm{tot}}}\sum_{j=0}^{M}\sum_{l\in\left\{ g,e\right\} }\frac{B_{l,j}^{2}\bigl(n-m\bigr)^{2}t_\mathrm{int}^{2}}{A_{l,j}}.
\end{align*}
The experimental $\mathcal{Q}_{\omega}$ is calculated by the coefficients
$A_{j}$ and $B_{j}$, which are obtained in the experiments of measuring
the virtual phase by varying $\varphi_{0}$. The results indicate
a sensitivity of measuring $p$ as
\[
\sigma_{p}\leq\frac{1}{\sqrt{\mathcal{Q}_{\omega}}}/\frac{\partial\omega}{\partial p}=\frac{1}{\sqrt{\mathcal{Q}_{\omega}}}\frac{1}{\chi}.
\]
For the direct implementation of the radiometry, the sensitivity of measuring $p$ is provided
as
\[
\sigma_{p}=\frac{1}{\sqrt{\mathcal{Q}_{p}}},
\]
with
\begin{align*}
\mathcal{Q}_{p}= & \frac{1}{t_{\mathrm{tot}}}\sum_{j=0}^{M}\sum_{l\in\left\{ g,e\right\} }\frac{1}{P_{l,j|M}}\left[\frac{\partial P_{l,j|M}}{\partial p}\right]^{2},
\end{align*}
where the probabilities $P_{l,j|M}$ and their slopes $\frac{\partial P_{l,j|M}}{\partial p}$ are obtained directly from experiments with an optimized $\varphi_0$.

\newpage{}

\end{document}